\documentclass[a4paper]{jpconf}
\usepackage{graphicx}
\usepackage{psfrag,fancyhdr,epsfig}

\usepackage{amssymb}
\usepackage{amsthm}
\usepackage{latexsym}
\newcommand{\be}{\begin{eqnarray}}
\newcommand{\ee}{\end{eqnarray}}
\newcommand{\bdm}{\begin{displaymath}}
\newcommand{\edm}{\end{displaymath}}

\begin{document}
\title{\textbf{Graviton Emission from a Schwarzschild Black Hole in the Presence of Extra Dimensions}}
\author{\textbf{O. Efthimiou}}
\date{}
\address{University of Ioannina, Physics Department\\
Theoretical Physics Division\\
45110, Ioannina,Greece} \ead{\textbf{me01445@cc.uoi.gr}}
 \vspace{1cm}
%\numberwithin{equation}{section}
%%%%%%%%%%%%%%%%%%%%%%%%%%%%%%%%%%%%%%%%%%%%%%%%%%%%%%%
\begin{abstract}
In this talk I presented a previously published work\footnote{
This work was done in collaboration with S. Creek, P. Kanti and K.
Tamvakis .} concerning
 the evaporation of $(4+n)$-dimensional non-rotating black holes into
 gravitons \cite{Creek}
. In this work we calculated the energy emission rate for
gravitons in the bulk obtaining analytical solutions of the master
equation satisfied by all three types (S,V,T) of gravitational
perturbations and presented graphs for the absorption probability
and the energy emission of the black hole in the bulk.
 \end{abstract}
%%%%%%%%%%%%%%%%%%%%%%%%%%%%%%%%%%%%%%%%%%%%%%%%%%%%%%%

\section{Introduction}
In the first part of the talk I gave a brief introduction to the
brane world scenario. In this higher-dimensional context, which
has received a lot of attention in the last few years \cite{ADD,
RS}, gravity propagates in $D=4+n$ dimensions ({\textit{Bulk}}),
while all Standard Model degrees of freedom are assumed to be
confined on a four-dimensional $D$-{\textit{Brane}}. Within this
scenario hierarchy problem can be solved and black holes can be
produced in
colliders or in cosmic ray interactions.\\
 In the last few years there has been considerable interest in theoretically studying these higher dimensional black holes  that may be created in near
future experiments ,perhaps even LHC (see \cite{Kanti,reviews} for
a review) . An important property of them is Hawking radiation.
According to it black holes are not completely black , but they
emit radiation in a thermal spectrum  characterized   by a
temperature $T_H$  such as a blackbody ,while they deviate from
the exact blackbody behavior by a factor called greybody factor.
For a spherically symmetric BH
$$
{{dN^{(s)} (\omega )} \over {dt}} = \sum\limits_\ell  {\sigma
^{(s)} _{\ell ,n} (\omega ){\omega  \over {\exp (\omega /T_H ) \pm
1}}{{d^{n + 3} k} \over {(2\pi )^{n + 3} }}}
$$
with $\sigma$ being the greybody factor which depends on
particle's energy, spin, angular momentum and on space
dimensionality. What is crucial, is that $\sigma$ has been proven
to be proportional to the incoming absorption probability, and the
following formulae is found to hold for a spherically symmetric
higher-dimensional black hole \be {{dE} \over {dt}} =
\sum\limits_\ell {N_\ell \left| {A_\ell } \right|^2 {\omega \over
{\exp (\omega /T_H ) \pm 1}}{{d\omega } \over {2\pi }}}
\label{emr} \ee giving the energy emission rate for a particle of
energy $\omega$ , with l being the angular momentum quantum number
, $A_l$ the absorption probability and $N_l$ a constant counting
the multiplicities of states with the same l.\\ This black hole
radiation has been the subject of both analytical and numerical
studies in the past. This includes lower spin degrees of freedom
\cite{kmr1, FS, HK1, BGK, Barrau, Jung} as well as graviton
emission in the bulk \cite{Naylor}. The study of graviton emission
in the bulk corresponds to the study of perturbations in a
gravitational background. Within this context , we studied  the
hawking radiation of a Schwarzschild Black hole through gravitons.
Gravitons emitted from a black hole will "see" the entire (4+n)
dimensional space-time (Bulk),will account for an energy loss from
the brane where the BH is located to the bulk, and their emission
rate will depend from
the dimensionality of space-time, namely n.\\
So what we will do is write the equations for graviton emission
,analytically solve the equations, compute the absorption
probability and from that compute the energy emission rate and its
dependence on n .
%%%%%%%%%%%%%%%%%%%%%%%%%%%%%%%%%%%%%%%%%%%%%%%%%%%%%%%
\section{Mathematical Background}
The gravitational background for a spherically symmetric , non
charged non rotating black hole was found to be \cite{TMP}
$$ ds^2  =  - f(r)dt^2  + {{dr^2 } \over
{f(r)}} + r^2 d\Omega ^2 _{2 + n}
$$
with
$$
f(r) = 1 - \left( {{{r_H } \over r}} \right)^{n + 1}
$$
and $d\Omega^2_{n+2}$ the line element in a (n+2) dimensional unit
sphere. The temperature of the black hole associated with Hawking
radiation is
$$
T_H  = {{n + 1} \over {4\pi r_H }}
$$
The equations that describe graviton emission from a Schwarzschild
black hole in (4+n) dimensions are known for a little while
\cite{IK} and the way to produce them is to perturb the metric
outside a black hole , use Einstein equations and thus get three
independent wave equations describing emitted gravitational waves
, analyzed  in terms of partial waves with specific angular
momentum.These equations are called scalar, vector and tensor
perturbation equation respectively and they describe the emitted
gravitons that correspond to each type of metric perturbation. All
three can be written in the form of a master equation for the
partial waves \be f{d \over {df}}\left( {f{{d\Phi } \over {dr}}}
\right) + (\omega ^2  - V)\Phi  = 0 \label{ph} \ee with V having
different form for every perturbation. For the case of tensor and
vector type we have: \be V_{T,V}  = \left[ {l(l + n + 1) + {{n(n +
2)} \over 4} - {{k(n + 2)^2 } \over 4}\left( {{{r_H } \over r}}
\right)^{n + 1} } \right] \label{V} \ee with l being the angular
momentum quantum number and k= -1 (3) for tensor (vector) type .
For the Scalar case the potential V is much more complicated but
the way to treat it is similar to the tensor-vector type, so we
will concentrate on the first case and
present the complete results for all three types in the end.\\
In order to solve the equations we will use an approximate method
, valid at the low energy limit of the equations. What we will do
is at first solve the equations close to the horizon $r_H$ , then
solve the equations far away from the horizon and in the end
stretch the two solutions and match them in the intermediate zone.
We must however take into consideration the boundary condition
that nothing can escape from the black hole after it crosses the
horizon, so in the solution we will have we must impose the
condition that there are no outgoing waves in the limit
r$\rightarrow r_H$
%%%%%%%%%%%%%%%%%%%%%%%%%%%%%%%%%%%%%%%%%%%%%%%%%%%%%%%

\section{Solving the Equations - Computing the Absorption Coefficient}
We will solve equation (\ref{ph}) for the potential (\ref{V}) ,
first at the near horizon regime. We change the variables from   r
to $f (r) = 1 - (r_H / r)^{n+1}$ ,and after making the field
redefinition
$$
\Phi  = f^\kappa  (1 - f)^\lambda  F
$$
and taking the limit  $r\rightarrow r_H$  ,the equation (\ref{ph})
can be brought to the following form \be f(1 - f){{d^2 F} \over
{df^2 }} + [c - (1 + a + b)]{{dF} \over {df}}] - abF = 0
\label{hg} \ee for the following choice of the parameters :

$$
a = \kappa  + \lambda  + {{(n + 2)} \over {2(n + 1)}} + G
\,\,\,\,\,\,\,\, b = \kappa  + \lambda  + {{(n + 2)} \over {2(n +
1)}} - G
$$
$$
c = 1 + 2\kappa\,\,\,\,\,\,\,\,\,\,\, G^{(T,V)}  = {{(1 + k)(n +
2)} \over {4(n + 1)}}
$$
$$
\lambda  = {1 \over {2(n + 1)}}\left[ { - 1 - \sqrt {(2l + n +
1)^2  - 4\omega ^2 r_H ^2 } } \right]\,\,\,\,\,\,\,\, \kappa  =  -
{{i\omega r_H } \over {n + 1}}
$$
Equation (\ref{hg}) has known solutions, the hypergeometric
functions , so the near horizon solution for tensor-vector mode
gravitons is  :
$$
\Phi _{NH}  = A_1 f^\kappa  (1 - f)^\lambda  F(a,b,c;f) + A_2 f^{
- \kappa } (1 - f)^\lambda  F(a - c + 1,b - c,2 - c;f)
$$
By expanding this solution in the limit $r\rightarrow r_H$ and
taking the boundary condition we get  $A_2=0$ ,and
$$
\Phi _{NH}  = A_1 f^\kappa  (1 - f)^\lambda  F(a,b,c;f)
$$
We then solve the equation in the far field regime , that is far
away from the horizon . The solution after taking $r>>r_H$ is
easily obtained in terms of the Bessel functions \be \Phi _{FF}  =
B_1 \sqrt r J_{l + (n + 1)/2} (\omega r) + B_2 \sqrt r Y_{l + (n +
1)/2} (\omega r) \label{ff} \ee
 We now have to match the two
solutions in the intermediate zone. So if we take the limit
$r>>r_H$ in the near horizon solution,the limit $\omega r<<1$ in
the far field solution and look in the low energy regime $\omega
r_H<<1$ , we see that the two solutions actually match and we can
compute the relation between the constants $B_1$ and $B_2$ of
equation (\ref{ff}). Thus we have constructed a solution for the
whole domain of r , valid for low energies , which we can use to
compute the absorption coefficient .We expand the far field
solution for $r\rightarrow\infty$ in terms of ingoing/outgoing
spherical waves, and we only need to compute the ratio of their
amplitudes. The result for the absorption coefficient put in a
compact form, is:
$$
\left| {A_l } \right|^2  = 4\pi \left( {{{\omega r_H } \over 2}}
\right)^{2l + n + 2} {{\Gamma \left( {l + {l \over {n + 1}} - G}
\right)^2 \Gamma \left( {l + {l \over {n + 1}} + G} \right)^2 }
\over {\Gamma \left( {l + {{n + 3} \over 2}} \right)^2 \Gamma
\left( {1 + {{2l} \over {n + 1}}} \right)^2 }}
$$
We found that the same formula holds for the scalar type as
well,with the only dependence from the type of perturbation
encoded in G , which is different for every type. In figure 1
there are plots of the absorption probability for all three types
of perturbation, for several combinations of n,$l$ .
%%%%%%%%%%%%%%%
\begin{figure}[t]
\begin{center}
\thicklines \mbox{
\includegraphics[height=5.6cm,clip]{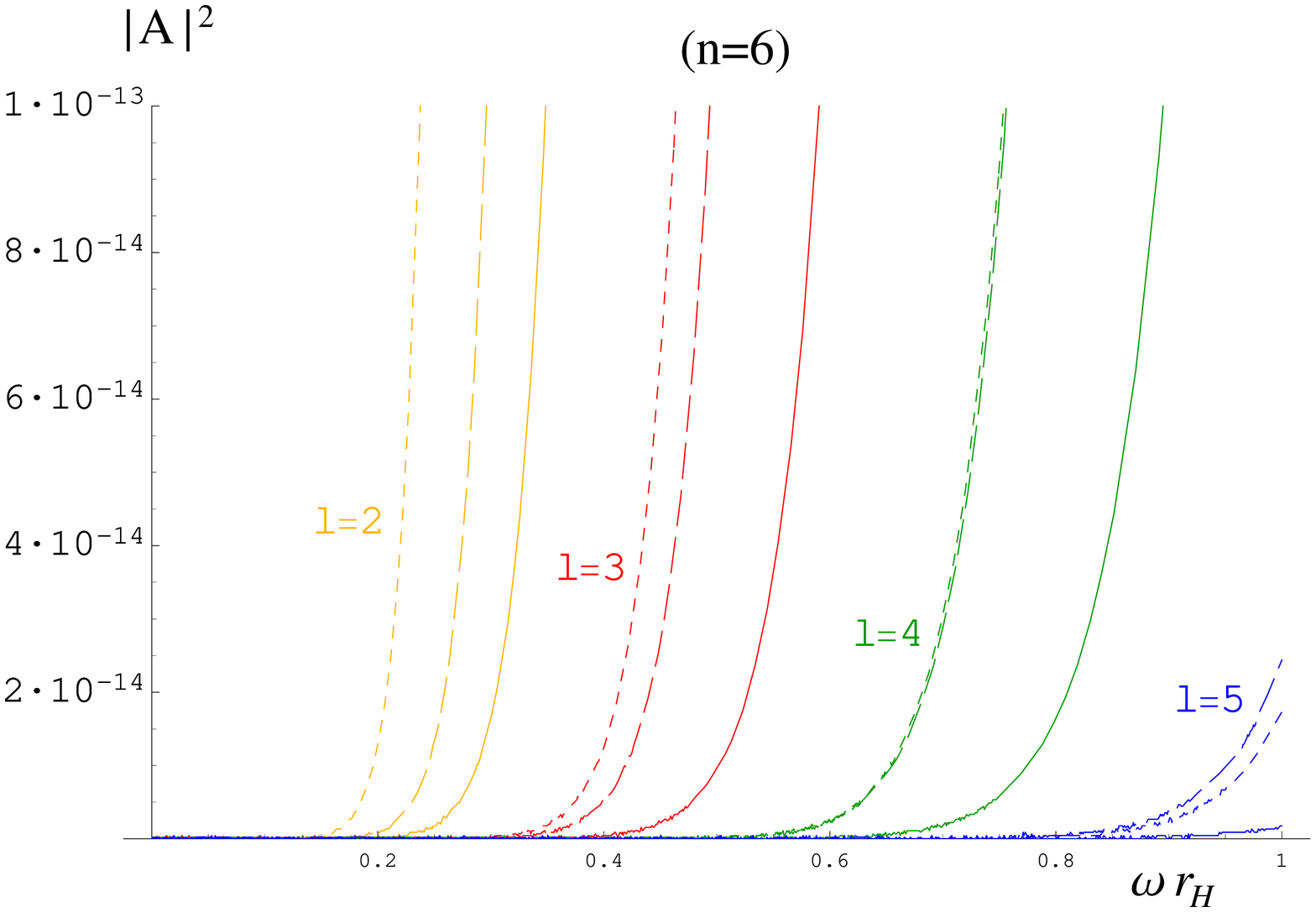}}\hspace*{-0.1cm}
{
\includegraphics[height=5.6cm,clip]{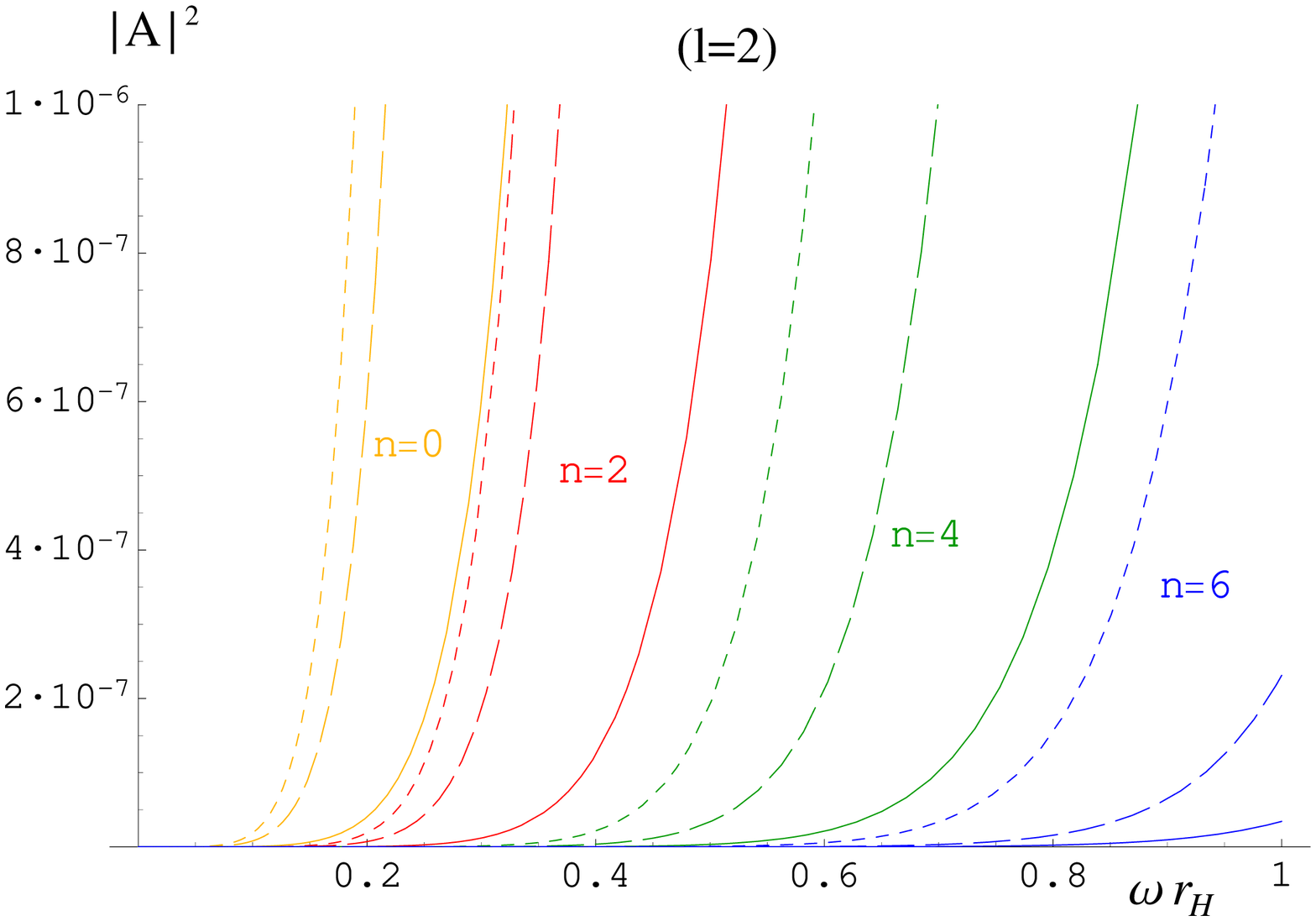}}
%\scalebox{0.8}{\rotatebox{0}{\includegraphics[width=\textwidth]{nefiopfig2.eps}}}
\caption{Absorption probability $|{\cal A}_l|^2$ for tensor (solid
lines), vector (short-dashed lines) and scalar (long-dashed lines)
gravitational perturbations in the bulk for: {\bf (a)} $n=6$, and
$l=2,3,4,5$, and {\bf (b)} $l=2$, and $n=0,2,4,6$.}
\label{gravitons-nl}
\end{center}
\end{figure}
%%%%%%%%%%%%%%
We can now proceed in computing the energy emission rates ,but
before that we have to know $N_l$ of equation (\ref{emr}) ,which
is the number counting the state multiplicities : there  is a
number of different states that correspond to the same angular
momentum number $l$ . This number depends on $l,n$  and
fortunately it has been computed in the literature for every type
of perturbation \cite{Rubin}
$$
N_\ell  ^{(T)}  = {{n(n + 3)(\ell  + n + 2)(\ell  - 1)(2\ell  + n
+ 1)(\ell  + n - 1)!} \over {2(\ell  + 1)!(n + 1)!}}
$$
$$
N_\ell  ^{(V)}  = {{(\ell  + n + 1)\ell (2\ell  + n + 1)(\ell  + n
- 1)!} \over {(\ell  + 1)!n!}} \,\,\,\,\,\, N_\ell  ^{(S)}  =
{{(2\ell + n + 1)(\ell  + n)!} \over {\ell !(n + 1)!}}
$$
Now we can compute the energy emission rates for all three types
and for every n and plot the results .
%%%%%%%%%%%%%%%
\begin{figure}[t]
\begin{center}
\thicklines \mbox{
\includegraphics[height=5.4cm,clip]{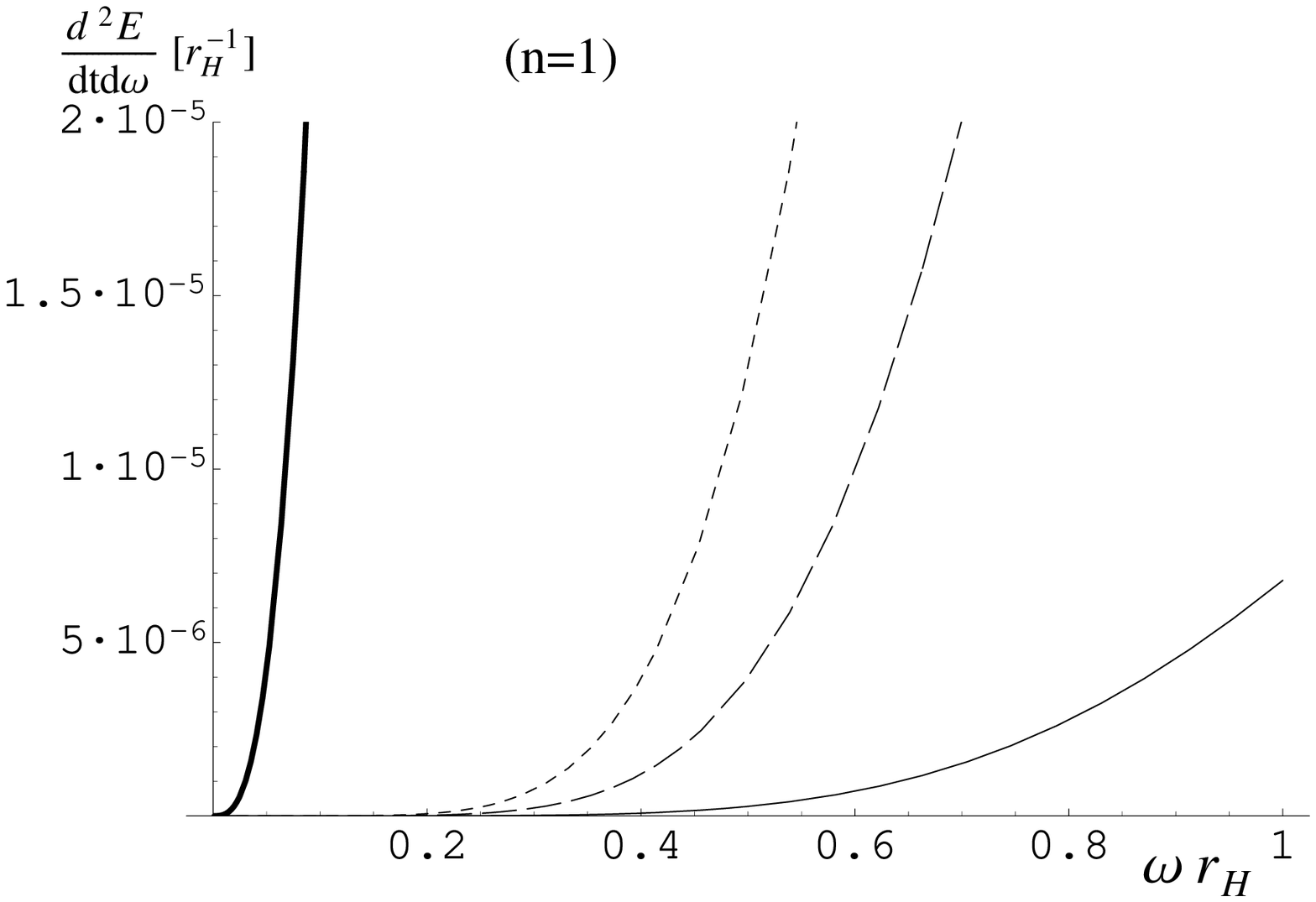}}\hspace*{-0.2cm}
{
\includegraphics[height=5.4cm,clip]{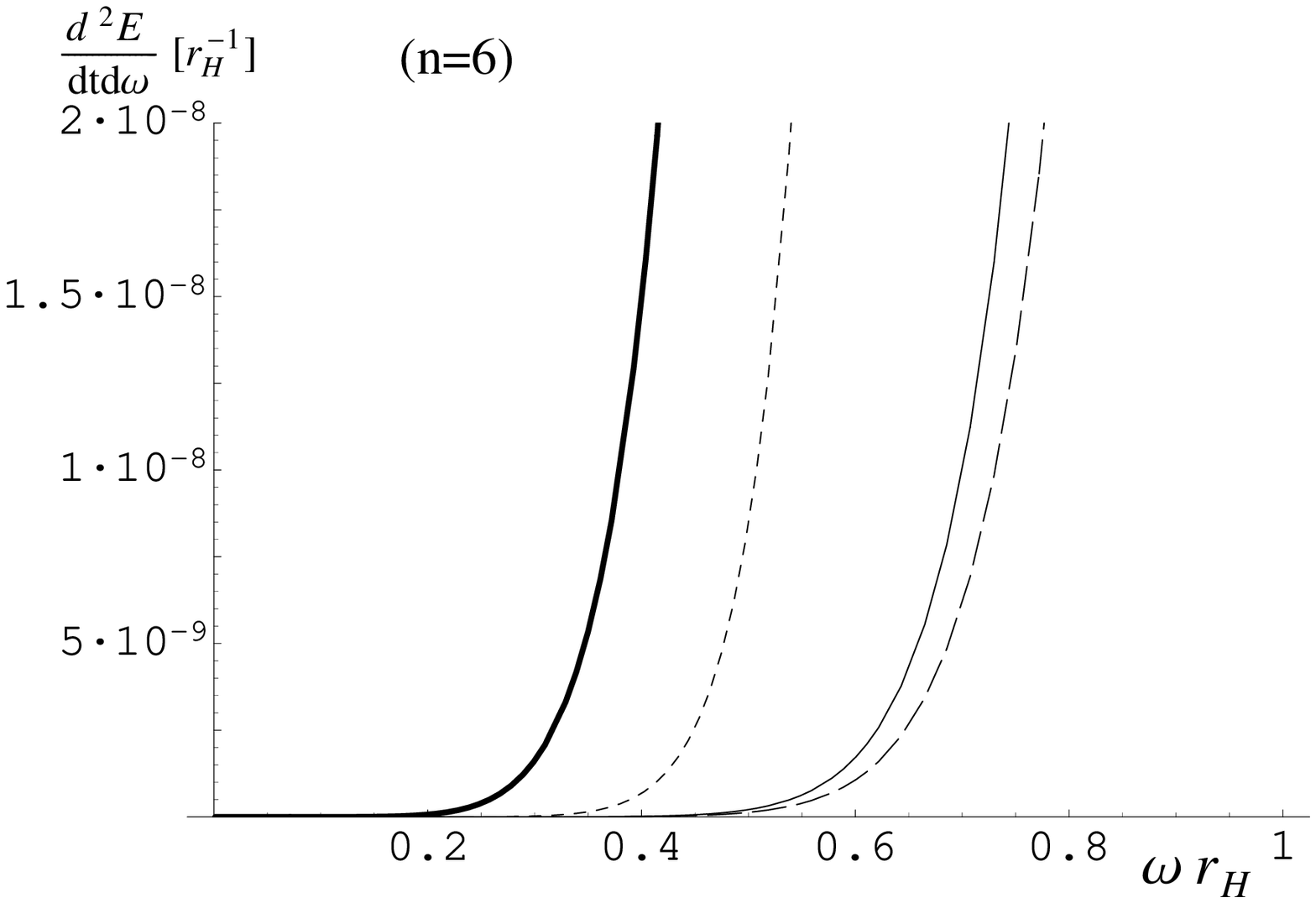}}
%\scalebox{0.8}{\rotatebox{0}{\includegraphics[width=\textwidth]{nefiopfig2.eps}}}
\caption{Energy emission rates for tensor (thin solid lines),
vector (short-dashed lines) and scalar (long-dashed lines)
gravitational perturbations, and scalar fields (thick solid lines)
in the bulk, for {\bf (a)} $n=1$, and {\bf (b)} $n=6$.}
\label{rates}
\end{center}
\end{figure}
%%%%%%%%%%%%%%
In figure 2 there are the energy emission rates for the case of 1
and 6 extra dimensions .\\
\section{Results-Discussion}
In this work I presented ,we studied graviton emission in the bulk
from a spherically symmetric (4+n) dimensional Schwarzschild black
hole.Our results, valid in the low-energy regime, are
complementary to existing studies in the intermediate energy
regime \cite{Naylor} and are in agreement with two papers
\cite{Park,Cardoso} that appeared in the literature while our work
was at its last stages. These results show that vector
perturbations are the dominant mode emitted in the bulk for all
values of $n$. The relative emission rates of the subdominant
scalar and tensor modes depend on $n$, with scalars foremost at
small $n$ and tensors more prevalent at high $n$. The absence of
the $l=0,1$ partial waves, dominant in the low-energy regime, from
all gravitational spectra causes even the total gravitational
emission rate to be subdominant to that of scalar fields in the
bulk. Finally, as previously found for bulk scalar fields
\cite{kmr1,HK1}, the energy emission rates for all types of
gravitational perturbations are suppressed with the number of
extra dimensions in the entire
low-energy regime.\\
The differential equations for the graviton emission were
analytically solved for low energies using a matching technique ,
according to which the equation solutions were found in the "near
horizon" and "far field" regime and were stretched and matched in
the intermediate zone. We thus computed the absorption probability
for all 3 types of perturbations  , and written down the equations
describing the black hole's Hawking radiation spectrum. Although
the radiation emitted in the bulk is not directly observable, it
determines the energy left for emission on our brane. In this
context, our results, in addition to their theoretical interest,
would be of particular use to experiments developed to detect the
low-energy spectrum of radiation emitted from a higher-dimensional
black hole

{\bf Acknowledgments.} O.E acknowledges I.K.Y. fellowship. This
research was co-funded by the European Union in the framework of
the Program $\Pi Y\Theta A\Gamma O PA\Sigma-II$ of the
{\textit{``Operational Program for Education and Initial
Vocational Training"}} ($E\Pi EAEK$) of the 3rd Community Support
Framework of the Hellenic Ministry of Education, funded by $25\%$
from national sources and by $75\%$ from the European Social Fund
(ESF).

\pagebreak

\end{document}